\documentclass{ws-ijmpd}
\usepackage[super,compress]{cite}

\begin{document}
\newcommand{\beq}[1]{\begin{equation}\label{#1}}
 \newcommand{\eeq}{\end{equation}}
 \newcommand{\bea}{\begin{eqnarray}}
 \newcommand{\eea}{\end{eqnarray}}

\markboth{Shuo Cao and Nan Liang}
{Interaction between dark energy and dark matter}

\title{Interaction between dark energy and dark matter: observational constraints from OHD, BAO, CMB and SNe Ia}

\author{Shuo Cao}
\address{Department of Astronomy, Beijing Normal University, Beijing 100875, China;\\
baodingcaoshuo@163.com}

\author{Nan Liang}
\address{Center for High Energy Physics, Peking University, Beijing 100871, China}

\maketitle

\begin{history}
\received{Day Month Year}
\revised{Day Month Year}
\end{history}

\begin{abstract}
In order to test if there is energy transfer between dark energy and
dark matter, we investigate cosmological constraints on two forms of
nontrivial interaction between the dark matter sector and the sector
responsible for the acceleration of the universe, in light of the newly revised
observations including OHD, CMB, BAO and SNe Ia.
More precisely, we find the same tendencies for both phenomenological forms of the interaction term $Q=3\gamma H\rho$, i.e., the parameter $\gamma$
to be a small number, $|\gamma|\approx 10^{-2}$. However, concerning the sign of the interaction parameter, we observe that $\gamma>0$ when the interaction between dark sectors is proportional to the energy density of dust matter, whereas the negative coupling ($\gamma<0$) is preferred by observations when the interaction term is proportional to dark energy density. We further discuss two possible explanations to this incompatibility and apply a quantitative criteria to judge the severity of the coincidence problem. Results suggest that the $\gamma_m$IDE model with a positive coupling may alleviate the coincidence problem, since its coincidence index $C$ is smaller than that for the $\gamma_d$IDE model, the interacting quintessence and phantom models by four orders of magnitude.

\end{abstract}


\ccode{PACS numbers:}

\section{Introduction}\label{sec1}
Many astrophysical and cosmological observations, such as Type Ia
Supernovae (SNe Ia) \cite{Riess98,Perlmutter99,Riess04,Knop07},
cosmic microwave background (CMB) from Wilkinson Microwave
Anisotropy Probe (WMAP)
\cite{Spergel03,Spergel07,Komatsu09,Komatsu10}, and the large scale
structure from Sloan Digital Sky Survey (SDSS)
\cite{Tegmark04,Eisenstein05} have indicated that the universe is
undergoing an accelerating expansion, which consequently leads to
the assumption of the existence of dark energy (DE), an exotic energy
with negative pressure and causes an accelerating expansion of our
universe at late times. The most simple candidate for these
uniformly distributed dark energy is considered to be in the form of
a cosmological constant ($\Lambda$) with a equation of state
$w=\rho_\Lambda/p_\Lambda\equiv-1$. However, the corresponding
$\Lambda$CDM model is always entangled with the coincidence problem:
The matter density $\rho_m$ decreases with the expansion of our
universe with $a^{-3}$ and the density of cosmological constant
$\rho_\Lambda$ does not change with the expansion of the universe,
whereas the dark energy density is comparable with the dark matter (DM) density today, why? Although many alternative models include the
scalar field models with dynamical EoS (e.g., the quintessence
\cite{Wetterich88,Ratra88,Caldwell98}, phantom
\cite{Caldwell02,Caldwell03}, k-essence \cite{Armendariz,Chiba02},
as well as quintom model \cite{Feng05,Feng06,Guo05a}), have been
proposed to alleviate the coincidence problem, the nature of dark
energy is still a mystery and the coincidence problem is still a puzzle.

It is natural to take into account a possible interaction between the dark matter and dark
energy through an interaction term. The interacting dark energy model (IDE) was first introduced to test the
coincidence problem, in which $\rho_m$ could decrease with the
expansion of our universe slower than $a^{-3}$. It should be stressed that interacting dark energy
scenarios have been studied by many authors. Amendola et al.(2000)
investigated a coupled quintessence (CQ) model by assuming an
exponential potential and a linear coupling \cite{Amendola00}, besides many other
background and perturbation constraints \cite{Bean08,LaVacca09,DeBernardis11}. On the other hand, it is
always assumed that dark energy and dust matter exchange energy
through an interaction term $Q$
 \bea
 &&\dot{\rho}_X+3H\left(\rho_X+p_X\right)=-Q, \nonumber\\
 &&\dot{\rho}_m+3H\rho_m=Q,\label{generalQ}
 \eea
which preserves the total energy conservation equation
$\dot{\rho}_{tot}+3H\left(\rho_{tot}+p_{tot}\right)=0$. The
interaction term $Q$ is extensively considered in the literature
~\cite{Dalal01,Chimento03,Setare04,Nojiri05,Szydlowski06,Bertolami07,Guo07,Chen10}.
In these IDE models with linear and nonlinear interactions in the dark sector,
the $\Lambda$CDM model without interaction between dark energy and dark matter is
characterized by $Q=0$, while $Q\neq0$ and denote non-standard cosmology.

In the previous works, various interacting DE models have been discussed with different astronomical observations, such as SNe Ia and limited number of Hubble parameters. Recently, the observational Hubble parameter data (OHD) have  become an effective probe both in cosmology and
astrophysics compared with SNe Ia, CMB and the baryonic acoustic oscillation (BAO) data, and it is more rewarding to apply OHD to
investigate the the properties of the dark energy directly. The reason is quite simple, it is obvious that these probes all use the distance
scale (e.g., the luminosity distance $d_L$, the shift parameter $R$, or the distance parameter $A$) measurement to determine cosmological
parameters, which needs the integrate of the Hubble parameter and therefore lose the fine structure and some more important
information of $H(z)$ \cite{Lin09}. However, the Hubble parameter depends on the differential age as a function of redshift $z$ in the form
$H(z)=-\frac{1}{1+z}\frac{dz}{dt}$, which provides a direct measurement for $H(z)$ through a determination of $dz/dt$. Jimenez et al.(2003) demonstrated the
feasibility of the method by applying it to a $z\sim 0$ sample \cite{Jimenez03}. By
using the differential ages of passively evolving galaxies
determined from the Gemini Deep Deep Survey (GDDS)~\cite{Abraham04}
and archival data~\cite{Treu01,Treu02,Nolan03a,Nolan03b},
Simon et al.(2005) determined 9 $H(z)$ data in the range $0\leq z \leq
1.8$ \cite{hz1}. These OHD data were also used to constrain
the parameters of cosmological models \cite{Samushia06,Yi07} and
some other relevant works \cite{Wei07a,Wei07b,Lazkoz07,Kurek08,Sen08}. For example, Wei \& Zhang(2007a) compared the 9 observational $H(z)$ data with some cosmological models with/without interaction between dark energy and dust matter and found that the OHD data with fairly large errors cannot severely constrain model parameters alone \cite{Wei07a}.

In this paper, following a phenomenological approach, we will explore the properties hidden in two particular kinds of interaction in the dark sector \cite{Wei07a,Wei07b,Zhang10}, and then confront the theoretical models with the newly available observational data. We extend the previous analysis by focusing on the newly compiled observational Hubble parameter data (OHD), together with the cosmic microwave background (CMB)
detected by the 9-year WMAP data \cite{Komatsu10}, the baryonic acoustic
oscillation (BAO) peak detected by large-scale correlation function
of luminous red galaxies from the Sloan Digital Sky Survey (SDSS) data release 7 (DR7)\cite{Padmanabhan12}, SDSS-III Baryon Oscillation Spectroscopic Survey (BOSS) \cite{Anderson12}, WiggleZ survey \cite{Blake12}, 6dFGS survey \cite{Beutler11}, and the newly revised Union2 SNe Ia data set \cite{Amanullah}.
This paper is organized as follows: In section~\ref{sec2}, we introduce the
observational data considered to test our theoretical models in this work, with the latter described in Section~\ref{sec3}. In section~\ref{sec4}, we derive two Hubble parameters in a flat universe and perform a Markov Chain Monte Carlo analysis spanning the full parameter space using different data sets. In Section~\ref{sec5}, we apply a quantitative criteria to judge the severity of the coincidence problem in the two models.
Finally, we summarize the main conclusions in Section~\ref{sec6}.

\section{Observational data}\label{sec2}

In order to probe the above models against observations, we consider four background tests which are directly related to the behavior of the function $H(z)$, i.e. the Hubble parameter as a function of the redshift, and present the results for different combined analyses of these four tests.

Besides the 9 observational OHD data introduced in the introduction, more recently, Stern et al.(2010) obtained the OHD data at 11
different redshifts detected from the differential ages of red-envelope galaxies \cite{hz2}. In fact, the value of the function $H(z)$ can be directly obtained from other astrophysical observations. For instance, \cite{Busca12} found it possible to determine a high precision measurement of the Hubble parameter at $z=2.3$ from observations of the BAO peak in the Ly $\alpha$ forest. Based on the above two techniques, a list of 28 independent $H(z)$ measurements have been compiled by Farooq \& Ratra(2013) to constrain cosmological parameters in three cosmological models \cite{Farooq13}, which will also be include into our analysis. The original data can be obtained in Refs.~\refcite{Blake12} and \refcite{hz1}-\refcite{Zhang12}. The statistical analysis is based on the $\chi^2$ function constructed as
\begin{equation}
\label{chi2H}
\chi^2_H=\sum_{i=1}^{28}\frac{[H(z_i)-H_{obs}(z_i)]^2}{\sigma_{hi}^2},
\end{equation}
where $H_{obs}(z_i)$ is the observational OHD data at the redshift $z_i$, $\sigma_{hi}$ is the $1\sigma$ uncertainty.

As is well known, the baryonic oscillations at recombination are expected to leave baryonic acoustic oscillations (BAO) in the power spectrum
of galaxies. The acoustic peak in the galaxy correlation function has now been detected over a range of redshifts from $z=0.1$
to $z=0.7$, which provides a standard ruler measuring the distance ratio
\begin{equation}
d_z=\frac{r_s(z_d)}{D_V(z_{\mathrm{BAO}})},
\end{equation}
where the distance scale $D_V$ is given by \cite{Eisenstein05}
\begin{equation} D_V(z_{\mathrm{BAO}})=\frac{1}{H_0}\big
[\frac{z_{\mathrm{BAO}}}{E(z_{\mathrm{BAO}})}\big(\int_0^{z_{\mathrm{BAO}}}\frac{dz}{E(z)}\big
)^2\big]^{1/3}~,
\end{equation}
and $r_s(z_d)$ is the comoving is the co-moving sound horizon scale at recombination redshift $z_d$, ie, $r_s(z_{\ast})
={H_0}^{-1}\int_{z_{\ast}}^{\infty}c_s(z)/E(z')dz'$.

We use six precise measurements of the BAO distance ratio from the Sloan Digital Sky Survey (SDSS) data release 7 (DR7) at $z=0.35$ \cite{Padmanabhan12} and SDSS-III Baryon Oscillation Spectroscopic Survey (BOSS) at $z=0.57$ \cite{Anderson12}, and by the clustering of WiggleZ survey \cite{Blake12} at $z=0.44, 0.60, 0.73$ and 6dFGS survey at $z=0.10$ \cite{Beutler11}. The best-fit values of these measurements are
\begin{eqnarray}
\hspace{-.5cm}\bar{\bf{P}}_{\rm{BAO}} &=& \left(\begin{array}{c}
{\bar d_{0.10}} \\
{\bar d_{0.35}} \\
{\bar d_{0.57}} \\
{\bar d_{0.44}} \\
{\bar d_{0.60}} \\
{\bar d_{0.73}}\\
\end{array}
  \right)=
  \left(\begin{array}{c}
  0.336\pm0.015\\
  0.113\pm0.002\\
  0.073\pm0.001\\
  0.0916\pm0.0071\\
  0.0726\pm0.0034\\
  0.0592\pm0.0032\\
\end{array}
  \right).
 \end{eqnarray}
The WMAP nine-year analysis (WMAP9) \cite{Hinshaw13} incorporate these BAO observations into a likelihood of the form and the corresponding $\chi^2$ reads
\begin{eqnarray}
\chi^2_{\mathrm{BAO}}=\Delta
\textbf{P}_{\mathrm{BAO}}^\mathrm{T}{\bf
C_{\mathrm{BAO}}}^{-1}\Delta\textbf{P}_{\mathrm{BAO}},
\end{eqnarray}
where ${\bf C_{\mathrm{BAO}}}^{-1}$ is the corresponding inverse covariance matrix.

The third test we use is the cosmic microwave background observations. The locations of peaks in the CMB temperature power spectrum in l-space depend on the co-moving scale of the sound horizon at recombination, and the angular distance to recombination. This is summarized by the data sets including the acoustic scale ($l_a$), the shift parameter ($R$), and the redshift of recombination ($z_{\ast}$). The acoustic scale can be expressed as
\begin{equation}
l_a=\pi\frac{\int_0^{z_{\ast}}\frac{dz'}{E(z')}/H_0}{r_s(z_{\ast})}.
\end{equation}
The so-called CMB shift parameter $R$ is related to the cosmology by
\begin{equation}
R=\Omega_{\mathrm{m}}^{1/2}\int_0^{z_{\ast}}\frac{dz'}{E(z')}.
\end{equation}
The redshift of recombination is
$z_{\ast}=1048[1+0.00124(\Omega_bh^2)^{-0.738}(1+g_{1}(\Omega_{\mathrm{m}}h^2)^{g_2})]$,
where $g_1$ and $g_2$ are related to the value of $\Omega_bh^2$ \cite{Hu96}.
From the WMAP9 measurements \cite{Hinshaw13}, the best-fit values of the data set are
\begin{eqnarray}
\hspace{-.5cm}\bar{\textbf{P}}_{\rm{CMB}} &=& \left(\begin{array}{c}
{\bar l_a} \\
{\bar R}\\
{\bar z_{\ast}}\end{array}
  \right)=
  \left(\begin{array}{c}
302.40\\
1.7246\\
1090.88\end{array}
  \right).
 \end{eqnarray}
The $\chi^2$ value of the CMB observation can be expressed as
\begin{eqnarray}
\chi^2_{\mathrm{CMB}}=\Delta
\textbf{P}_{\mathrm{CMB}}^\mathrm{T}{\bf
C_{\mathrm{CMB}}}^{-1}\Delta\textbf{P}_{\mathrm{CMB}},
\end{eqnarray}
where $\Delta\bf{P_{\mathrm{CMB}}} =
\bf{P_{\mathrm{CMB}}}-\bf{\bar{P}_{\mathrm{CMB}}}$, and ${\bf
C_{\mathrm{CMB}}}^{-1}$ is the corresponding inverse covariance matrix \cite{Hinshaw13}.
Notice that $d_z$, $l_a$ and $R$ are independent of $H_0$, thus these quantities can provide robust constraint as complement to OHD on
dark energy models.

The fourth test comes from the supernova type Ia (SNIa) test. It is commonly believed that SNe Ia all have the same intrinsic
luminosity, and thus can be used as ``standard candles''. Recently, the Supernova Cosmology Project (SCP) collaboration has released
their Union2 compilation which consists of 557 SNe Ia \cite{Amanullah}. The Union2 compilation is the largest published
and spectroscopically confirmed SNe Ia sample to date, which are used in this paper. In the calculation of the likelihood from SNe Ia
we have marginalized over the nuisance parameter
\begin{equation}
\label{chi2SN}
\chi^2_{SN}=A-\frac{B^2}{C}+\ln\left(\frac{C}{2\pi}\right),
\end{equation}
where $A=\sum_i^{557}{(\mu^{\rm data}-\mu^{\rm
th})^2}/{\sigma^2_i}~, B=\sum_i^{557}{\mu^{\rm data}-\mu^{\rm
th}}/{\sigma^2_i}~, C=\sum_i^{557}{1}/{\sigma^2_i}$, $\mu^{\rm
data}$ is the distance modulus obtained from observations and
$\sigma_i$ is the total uncertainty of SNe Ia data.

\begin{table}[ph]
\tbl{The best-fit value of parameters \{$w_X$, $\gamma_m$ ($\gamma_d$), $\Omega_{m0}$, and $H_0$\} for the two interacting dark energy models with 1-$\sigma$ and 2-$\sigma$ uncertainty for OHD, OHD+BAO+CMB, and OHD+SNe+BAO+CMB, respectively.}
{{\scriptsize
 \begin{tabular}{|c|c|c|c|} \hline\hline
                          &     OHD               &     OHD+BAO+CMB                     &     OHD+SNe+BAO+CMB        \\ \hline
 & \multicolumn{3}{c|}{The $\gamma_m$ IDE Model}  \\
  \cline{2-4}
 $w_X$                  \ \ & \ \ $-1.7563_{-0.5318(-0.7650)}^{+0.4753(+0.6435)}$\ \  & \ \ $-0.5184_{-0.2695(-0.4269)}^{+0.1415(+0.1829)}$\ \  & \ \ $-0.9378_{-0.1619(-0.2256)}^{+0.1726(+0.2385)}$\ \ \\
 $\gamma_m$             \ \ & \ \ $0.0102_{-0.0083(-0.0107)}^{+0.0585(+0.1101)}$\ \  & \ \ $0.0191_{ -0.0100(-0.0130)}^{+0.0134(+0.0209)}$\ \  & \ \ $0.0073_{-0.0034(-0.0047)}^{+0.0048(+0.0071)}$\ \ \\
 $\Omega_{m}$          \ \ & \ \ $0.1425_{-0.0382(-0.0427)}^{+0.0636(+0.0993)}$\ \  & \ \ $0.3664_{-0.0519(-0.0779)}^{+0.0415(+0.0561)}$\ \   & \ \ $0.2940_{-0.0268(-0.0368)}^{+0.0287(+0.0417)}$\ \ \\
 $H_0/100$             \ \ & \ \ $0.8698_{-0.1476(-0.1967)}^{+0.1301(+0.1301)}$\ \  & \ \ $0.5206_{-0.0535(-0.0701)}^{+0.0857(+0.1330)}$\ \  & \ \ $0.6483_{-0.0503(-0.0698)}^{+0.0430(+0.0625)}$\ \ \\ \hline

 & \multicolumn{3}{c|}{The $\gamma_d$ IDE Model}  \\
 \cline {2-4}

 $w_X$                  \ \ & \ \ $-1.5342_{-0.5203(-0.6585)}^{+0.3910(+0.4779)}$\ \  & \ \ $-1.3890_{-0.3145(-0.4381)}^{+0.2731(+0.3787)}$\ \  & \ \ $-1.1502_{-0.1641(-0.2340)}^{+0.1521(+0.2106)}$\ \ \\
 $\gamma_d$             \ \ & \ \ $-0.2091_{-0.0748(-0.1067)}^{+0.1211(+0.2248)}$\ \  & \ \ $-0.0195_{-0.0177(-0.0239)}^{+0.0145(+0.0208)}$\ \  & \ \ $-0.0137_{-0.0153(-0.0226)}^{+0.0147(+0.0200)}$\ \ \\
 $\Omega_{m}$          \ \ & \ \ $0.0203_{-0.0043(-0.0049)}^{+0.0767(+0.1314)}$\ \  & \ \ $0.2390_{-0.0340(-0.0476)}^{+0.0355(+0.0527)}$\ \   & \ \ $0.2690_{-0.0232(-0.0308)}^{+0.0242(+ 0.0332)}$\ \ \\
 $H_0/100$             \ \ & \ \ $0.8841_{-0.1263(-0.1604)}^{+0.1158(+0.1158)}$\ \  & \ \ $0.7716_{-0.0563(-0.0798)}^{+0.0653(+0.0957)}$\ \  & \ \ $0.7197_{-0.0316(-0.0448)}^{+0.0341(+0.0487)}$\ \ \\ \hline\hline
 \end{tabular}} \label{constraint}}
\end{table}

\section{The interacting dark energy model}\label{sec3}

In a flat FRW metric we consider the universe composed by pressureless matter, $\rho_m$, and a dark energy component $\rho_X$.
The Friedmann equation in this case becomes
\begin{equation}
{H^2}\equiv \pi G(\rho_m+\rho_X)/3
\end{equation}
Allowing the components to interact through an interaction term $Q$, which can be an arbitrary function of the Hubble parameter $H$
and the energy densities of dust matter $\rho_m$ and dark energy $\rho_X$, we can study the energy transfer between DM and DE.

Next we need to identify the specific form of $Q$. In this paper, we will focus a simple assumption \bea Q=3\gamma H\rho, \eea where $\gamma$ is a
constant parameter quantifying the extent of interaction between dust matter and dark energy. We assume
that the EoS of dark energy $w_X\equiv p_X/\rho_X$ is a constant in
spatially flat FRW metric. When working out the value of $\gamma$,
we can see the extent of interaction and transfer direction between
dark energy and dark matter. For $\gamma<0$, the energy is transferred
from dark matter to dark energy; while for $\gamma>0$, the energy is
transferred from dark energy to dark matter, and the coincidence problem can be alleviated. In the literature, two choices of $Q$ have been considered: $Q=3\gamma_m H\rho_m$ and $Q=3\gamma_d H\rho_X$. Both of two cases will be analyzed in the subsequent section and the model parameters are determined by applying the maximum
likelihood method of $\chi^{2}$ fit by using the Markov Chain Monte Carlo (MCMC) method \cite{Lewis02}.

\section{The constraint results}\label{sec4}

In this section we will place some constraints on the two IDE models, using the observational Hubble $H(z)$ data (OHD) and the Union2 compilation of SNe Ia.
Following the method of Ref.~\refcite{Farooq13}, we firstly use the 28 independent OHD data points to constrain cosmological model parameters and then
present the results for a combined analysis of other observational tests.
Basically, the model parameters are determined by minimizing
\begin{equation}
\label{chi2min} \chi^2=\chi^2_{H}(+\chi^2_{\rm SN}+\chi^2_{\rm BAO}+\chi^2_{\rm CMB}).
\end{equation}
Thus, we obtain the best-fit values for six cases and calculate the corresponding marginal $1\sigma$ and $2\sigma$ error bars as
it can be seen in Table~\ref{constraint}. \\

\subsection{The $\gamma_m$ IDE Model}

We firstly consider the interaction term expresses as $Q=3\gamma_m H\rho_m$.
The analytical derivation of this model was presented in Ref.~\refcite{Wei07a} and here we only display the result.
In spatially flat FRW metric, for the $\gamma_m$ IDE model with a constant EoS of
dark energy $w_X$, the Friedmann equation is
\begin{eqnarray}
{E^2(z)}&=&\frac{w_X
\Omega_{m}}{\gamma_m+w_X}(1+z)^{3(1-\gamma_m)}+\left(1-\frac{w_X
\Omega_{m}}{\gamma_m+w_X}\right)(1+z)^{3(1+w_X)} \,.
\end{eqnarray}

\begin{center}
 \begin{figure}[htbp]
\centering
  \includegraphics[width=0.5\textwidth]{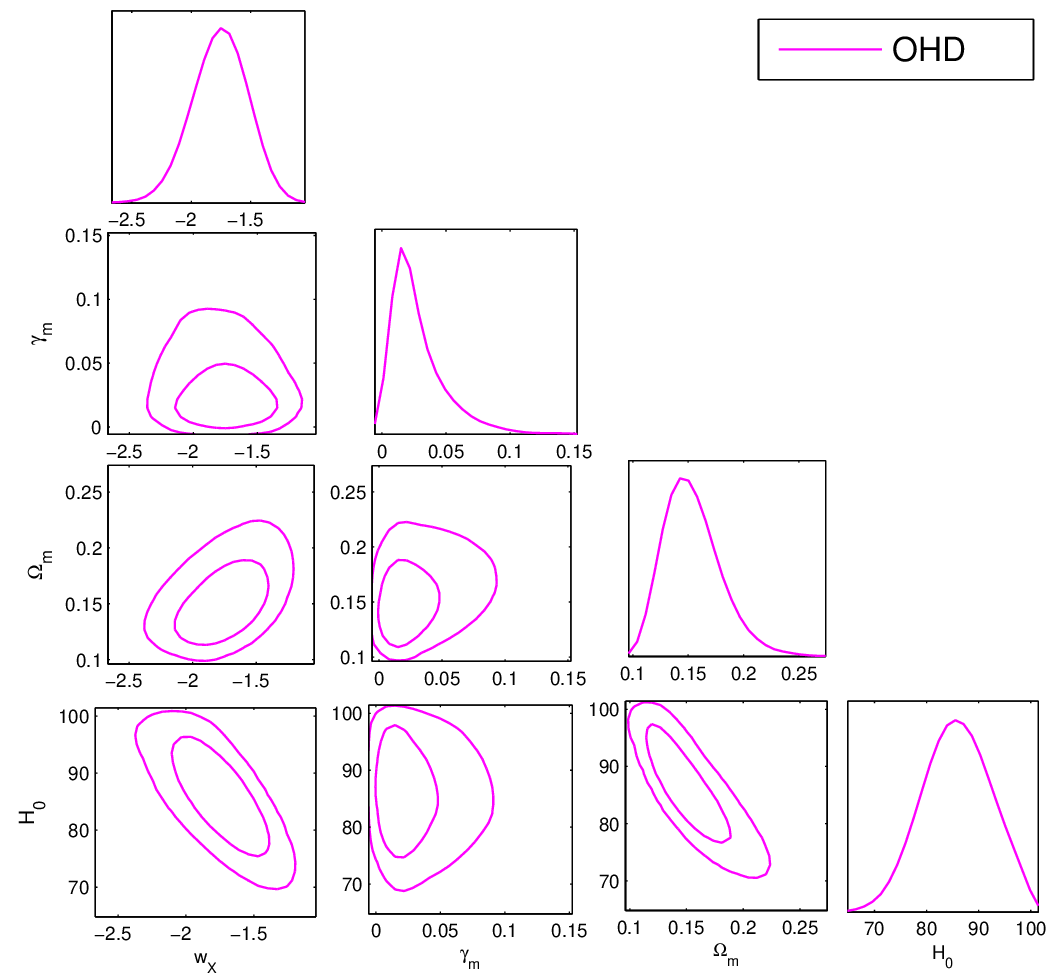}\includegraphics[width=0.5\textwidth]{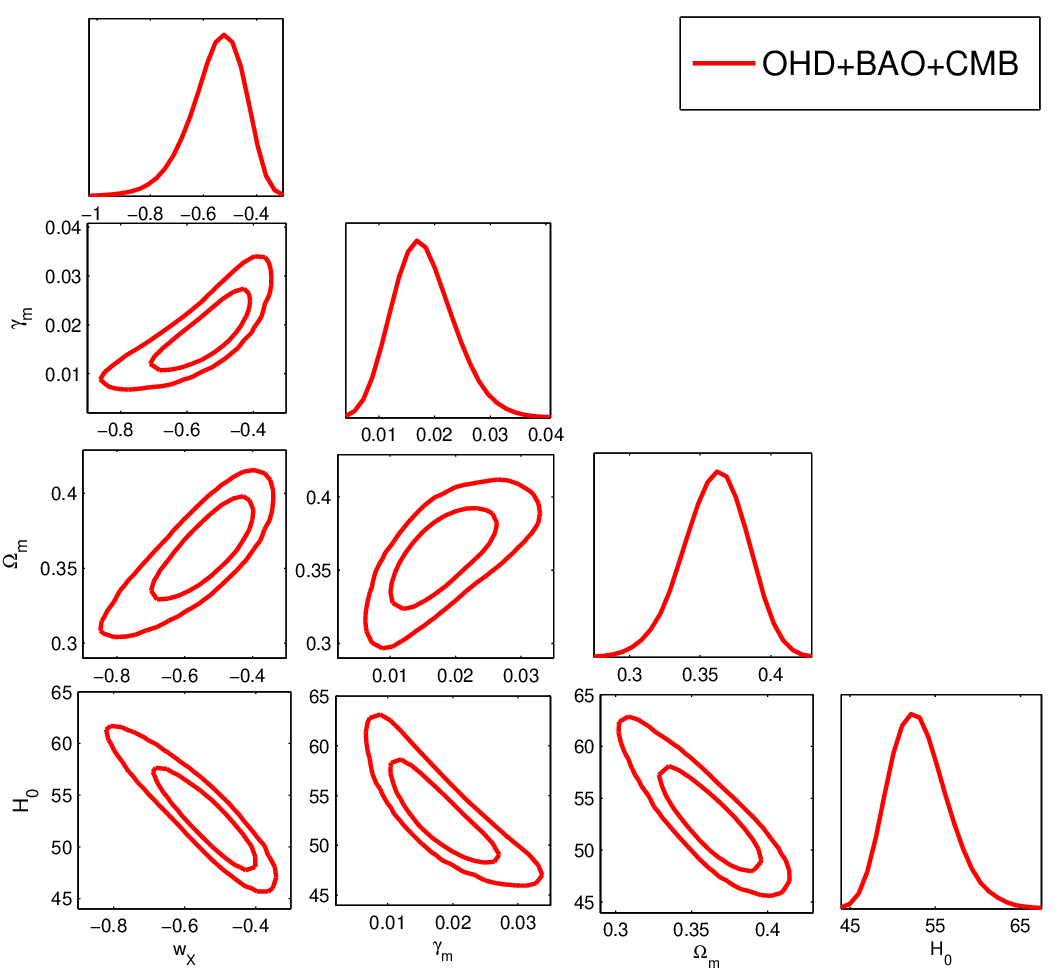}
 \caption{\label{1.1} The 2-D regions and 1-D marginalized distribution with the
1-$\sigma$ and 2-$\sigma$ contours of parameters $w_X$, $\gamma_m$, $\Omega_{\rm m}$, and $H_0$ in the $\gamma_m$IDE model, for the OHD data and the combined data set OHD+BAO+CMB.}
 \end{figure}
 \end{center}

\begin{center}
 \begin{figure}[htbp]
\centering
 \includegraphics[width=0.6\textwidth]{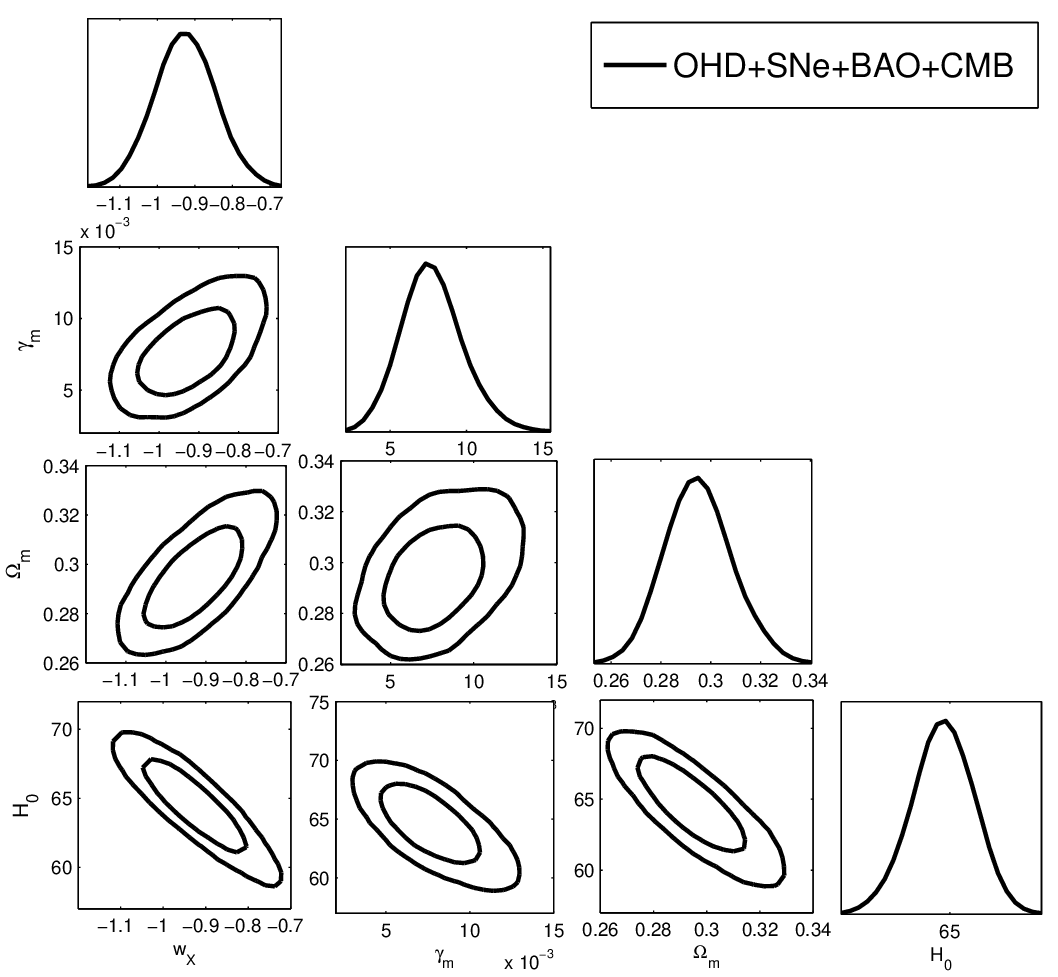}
 \caption{\label{1.2} The same as Fig.~\ref{1.1}, but for the combined data set OHD+SNe+BAO+CMB.}
 \end{figure}
 \end{center}

The 1-D probability distribution of each parameter ($w_X$, $\gamma_m$, $\Omega_{\rm m}$, and $H_0$) and 2-D plots for parameters between
each other, with different observational data sets (OHD, OHD+BAO+CMB, and OHD+SNe+BAO+CMB) are shown in Fig.~\ref{1.1}-\ref{1.2}.

Fig.~\ref{1.1} shows the contours constrained from the OHD only, with the values of the parameters ($w_X=-1.7563_{-0.5318(-0.7650)}^{+0.4753(+0.6435)}$,
$\gamma_m=0.0102_{-0.0083(-0.0107)}^{+0.0585(+0.1101)}$, and $\Omega_{m}=0.1425_{-0.0382(-0.0427)}^{+0.0636(+0.0993)}$) within 68.3\% confidence level.
With OHD+BAO+CMB, the best-fit value for the parameters are $w_X=-1.3890_{-0.3145}^{+0.2731}$, $\Omega_m=0.3664_{-0.0519}^{+0.0415}$, and $\gamma_m=0.0191_{ -0.0100}^{+0.0134}$. Compared to Fig.~\ref{1.1}, the allowed region of $\gamma_m$ is remarkably reduced in Fig.~\ref{1.2}.
In order to obtain more precise constraints, we also choose to include the Union2 compilation of SNe Ia data. Quantitatively, the value of
the interaction term $\gamma_m$ varies over the interval [0.004,0.012] within $1\sigma$ error region ($\gamma_m=0.0073_{-0.0034}^{+0.0048}$) whereas $w_X\in [-0.94,-1.10]$ ($w_X=-0.9378_{-0.1619}^{+0.1726}$). The present dust matter density parameter ranges from 0.27 to 0.32 ($\Omega_m=0.2940_{-0.0268}^{+0.0287}$) at 68.3\% C.L., and the constraint result on the Hubble parameter $h=0.6483_{-0.0503}^{+0.0430}$ is consistent with the observations obtained from 28 independently calibrated Cepheids and the distant, Cepheid-calibrated SNe Ia \cite{Tammann08}.

It is obvious that the OHD data, when combined to CMB and BAO observations, can give more stringent
constraints on this phenomenological interacting scenario. Moreover, the global statistical analysis with the Union2 data set may also significantly reduce the statistical errors of the four parameters, especially the interaction term $\gamma_m$.
From the results, the parameter $\gamma_m$, which stands for the interaction between DE and DM, seems to be vanishing or slightly larger than 0, which
means it is possible to have positive coupling between DE and DM. In addition, the constraining results in this work with the joint observational data are more stringent than previous results for constraining IDE model parameters
with other combined observations, for example, the 182 Gold SNe Ia samples, the shift parameter of CMB given by the WMAP3 observations, the BAO measurement from the Sloan Digital Sky Survey, and the age estimates of 35 galaxies \cite{Feng08}. \\

\subsection{The $\gamma_d$ IDE Model}

As mentioned in the introduction, we are interested to seek a cosmological model with the interaction term related to different forms of matter density (dust matter or dark energy), e.g., the interaction term is proportional to the matter density of dark energy
\begin{equation}
Q=3\gamma_d H\rho_X,
\label{gammad}
\end{equation}
where $\gamma_d$ corresponds to the constant quantifying the extent of interaction. For the convenience of numerical computing, in the spatially flat FRW metric, we introduce the so-called $e$-folding time $N\equiv\ln a=-\ln (1+z)$, $\tilde{\rho}_m\equiv 8\pi G\rho_m/(3H_0^2)$ and $\tilde{\rho}_X\equiv 8\pi G\rho_X/(3H_0^2)$.

With such kind of simplification, the Hubble parameter reads
\begin{eqnarray}
{H^2}&\equiv& \pi G(\rho_m+\rho_X)/3 \nonumber\\
     &=&H_0^2(\tilde{\rho}_X+\tilde{\rho}_m) \label{H2}
\end{eqnarray}
and combining with Eq.~(\ref{gammad}), the corresponding Eq.~(\ref{generalQ}) becomes
 \bea
 &&\frac{d\tilde{\rho}_X}{dN}=-3\tilde{\rho}_X
 (1+w_X)-3\gamma_d\tilde{\rho}_X \label{rhox}\\
 &&\frac{d\tilde{\rho}_m}{dN}=-3\tilde{\rho}_m+3\gamma_d\tilde{\rho}_X. \label{rhom}
 \eea
where $w_X$ is the time-independent EoS of dark energy. Considering the initial condition for integrating Eq.~(\ref{rhox}): $\tilde{\rho}_X(N=0)=1-\Omega_{m}$, where $\Omega_m \equiv 8\pi G \rho_{m0}/(3H_0^2)$ is the present fractional energy density of the dust matter, the Hubble parameter for this interacting dark energy model (Eq.~(\ref{H2})) expresses as
\begin{eqnarray}
{E^2(z)}&\equiv&(H/H_0)^2 \nonumber\\
        &=&(1-\Omega_{m})(1+z)^{3(1+\gamma_d+w_X)}\nonumber\\
        &+&\frac{w_X\Omega_{m}+\gamma_d+\gamma_d(\Omega_{m}-1)(1+z)^{3(\gamma_d+w_X)}}{w_X+\gamma_d}(1+z)^{3}
\end{eqnarray}

\begin{center}
 \begin{figure}[htbp]
\centering
 \includegraphics[width=0.5\textwidth]{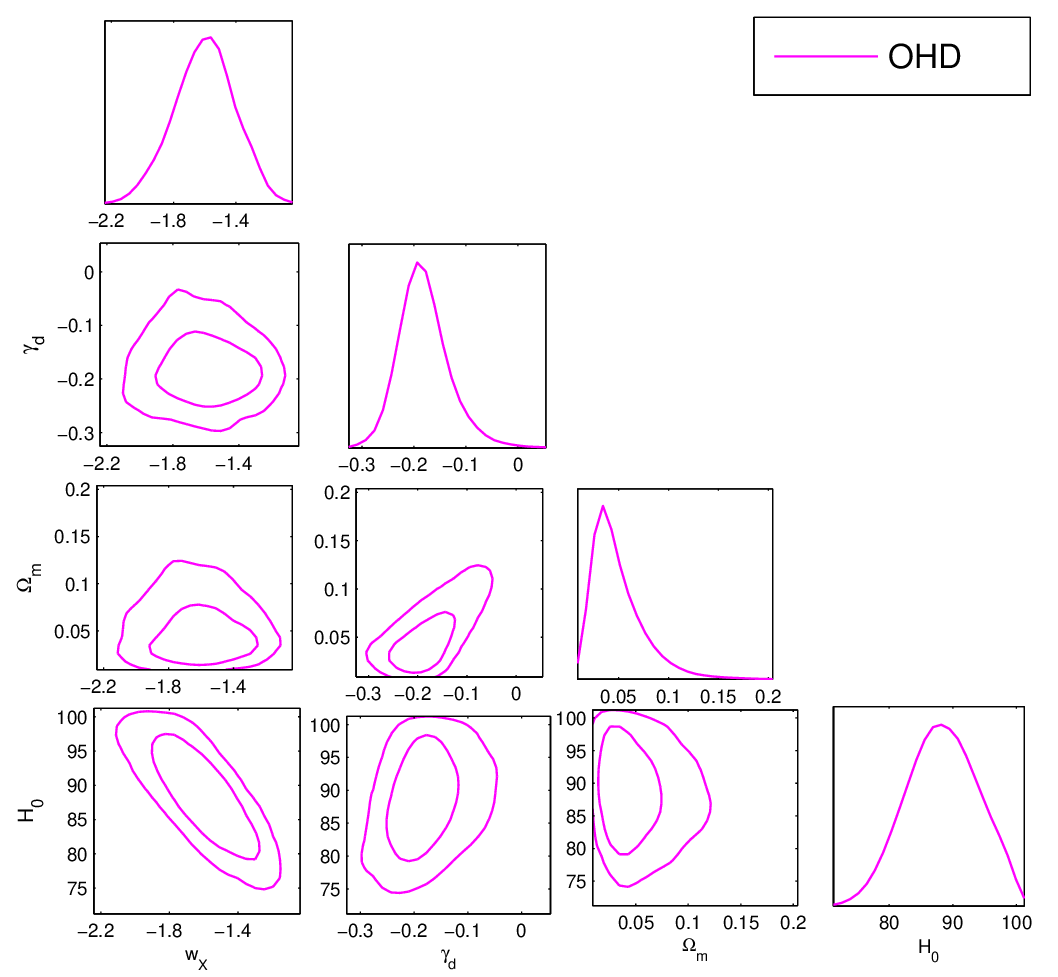}\includegraphics[width=0.5\textwidth]{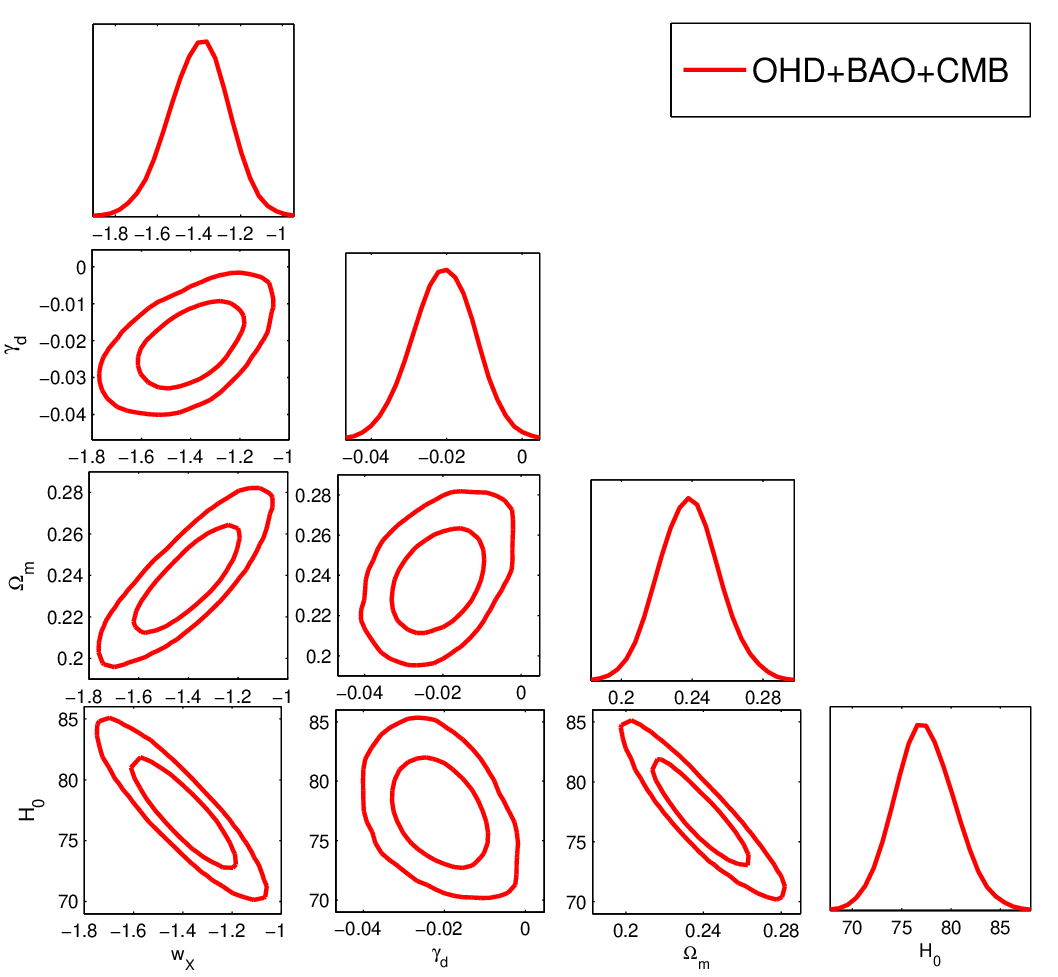}
 \caption{\label{2.1} The 2-D regions and 1-D marginalized distribution with the
1-$\sigma$ and 2-$\sigma$ contours of parameters $w_X$, $\gamma_d$, $\Omega_{\rm m}$, and $H_0$ in the $\gamma_d$IDE model, for the OHD data and the combined data set OHD+BAO+CMB.}
 \end{figure}
 \end{center}

\begin{center}
 \begin{figure}[htbp]
\centering
 \includegraphics[width=0.6\textwidth]{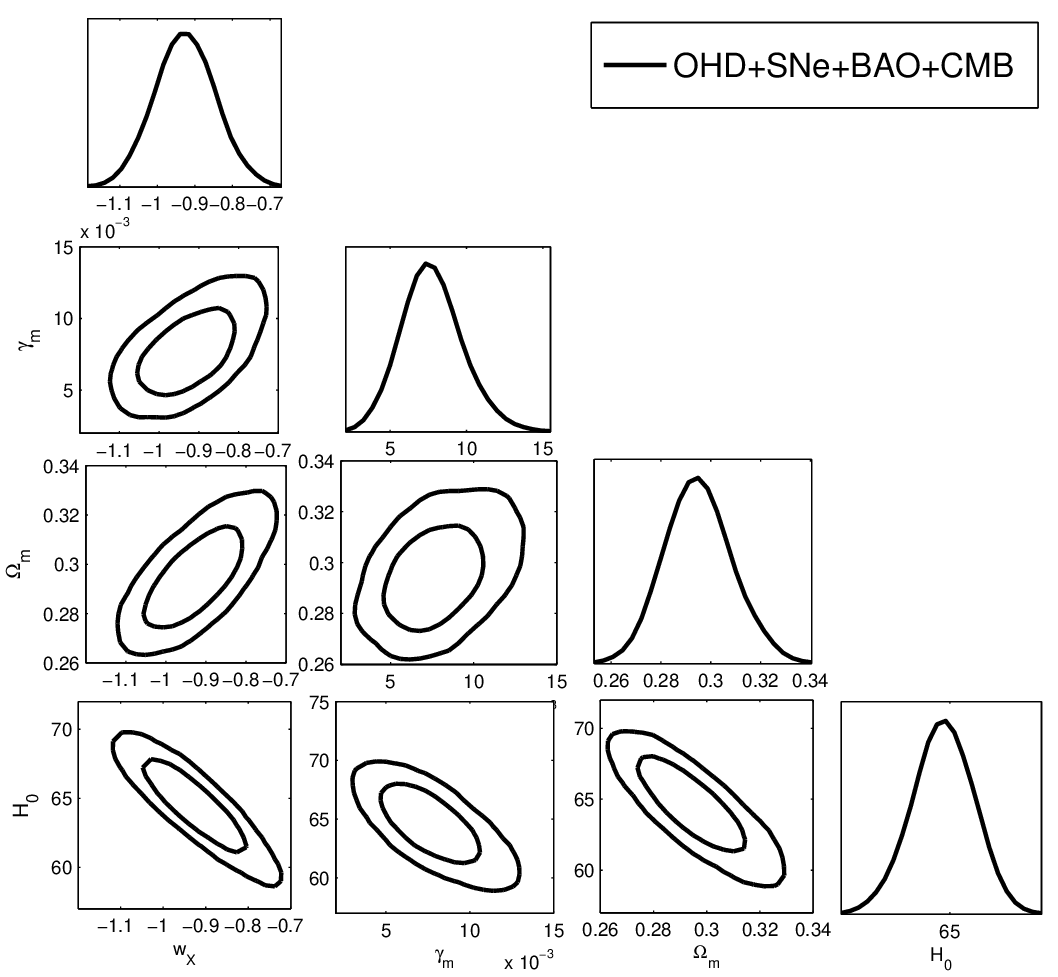}
 \caption{\label{2.2} The same as Fig.~\ref{2.1}, but for the combined data set OHD+SNe+BAO+CMB.}
 \end{figure}
 \end{center}

The joint confidence regions with different observational data sets (OHD, OHD+BAO+CMB and OHD+SNe+BAO+CMB) for the $\gamma_d$ IDE model
are showed in Fig.~\ref{2.1}-\ref{2.2}. We also explicitly present the best-fit values of
parameters with 1-$\sigma$ and 2-$\sigma$ uncertainties for the
$\gamma_d$ IDE model in Table~\ref{constraint}.

By minimizing the corresponding $\chi^2$ with the OHD data, we find that the best-fit values for the main parameters are
$w_X=-1.5342_{-0.5203(-0.6585)}^{+0.3910(+0.4779)}$, $\gamma_d=-0.2091_{-0.0748(-0.1067)}^{+0.1211(+0.2248)}$, and $\Omega_m=0.0203_{-0.0043(-0.0049)}^{+0.0767(+0.1314)}$. Notice that the matter
density $\Omega_m$ is weakly constrained with the observational $H(z)$ data only. However, CMB and BAO data, which are
taken as priors in our treatment and combined with other data, can help test the constraining power of OHD \cite{Gong12}.
The combined analysis with OHD+BAO+CMB provides the best-fit parameters as $w_X=-1.3890_{-0.3145}^{+0.2731}$, $\gamma_d=-0.0195_{-0.0177}^{+0.0145}$, and $\Omega_m=0.2390_{-0.0340}^{+0.0355}$. In order to include the Union2 data into constraint, we also perform a global statistical analysis with OHD+SNe+BAO+CMB by taking into account a global minimization of the four parameters. The latter procedure leads to the best-fit values $(w_X, \gamma_d, \Omega_m)=(-1.1502_{-0.1641}^{+0.1521},-0.0137_{-0.0153}^{+0.0147},0.2690_{-0.0232}^{+0.0242})$, and the constraint result on the Hubble parameter $h=0.7197_{-0.0316}^{+0.0341}$ is in qualitative agreement with the final results of the Hubble Space Telescope (\textit{HST}) key project that measured the Hubble constant $H_0=72\pm8 \rm{kms}^{-1}\rm{Mpc}^{-1}$ \cite{Freedman01}.

From Fig.~\ref{1.1}-\ref{2.2} and Table~\ref{constraint}, it is obvious that the constraints on this interacting scenario favor $\gamma<0$, suggesting that the energy is transferred from dark matter to dark energy and the coincidence problem is quite severe, a result
consistent with the previous results by using other observational data including the 182 Gold SNe Ia and 397 Constitution SNe Ia
samples \cite{Chen10,Feng07}.

Now it is worthwhile commenting on the results in the light of the recent observational data including the 9 years WMAP results.
We find the same tendencies for both phenomenological forms of interaction with two parameterizations of $Q$, i.e., the parameter $\gamma$
to be a small number, $|\gamma|\approx 10^{-2}$. The small coupling result is consistent with that independently obtained by galaxy cluster analysis \cite{Abdalla07} and combined analysis including the look back time data \cite{Feng08}. However, the difference between the constraint results made with the two IDE models is also most sharpest concerning the sign of the interaction parameter. We observe that $\gamma>0$ when the interaction between dark sectors is proportional to the energy density of dust matter, whereas the negative coupling ($\gamma<0$) is preferred by the latest observations when the interaction term is proportional to the DE energy density. In fact, the positive coupling is required to alleviate the coincidence problem, which is also the requirement of the second law of thermodynamics \cite{Pavon07}.

The first possible solution to this incompatibility is the proportional relation between the values of the EoS parameter and the interaction term $\gamma_d$ ($\gamma_m$), which can be seen from Fig.~\ref{1.1}-\ref{2.1}. Besides, the actual values of the equations of state of dark energy are quite different for the two phenomenological forms of interaction. For the $\gamma_d$IDE model, the best-fit value we obtain for the EoS parameter is $w_X=-1.1502_{-0.1641}^{+0.1521}$. Under this criteria, this model seems to behave in the same way as a quintom dark energy model. However, the $\gamma_m$IDE model exhibits a quintessence behavior and it has more possibility for the EoS parameter to cross quintessence divide line $w_X=-1$ at $1\sigma$ C.L. The other possible explanation to this discrepancy is the value of $\Omega_m$, compared with the $\gamma_m$IDE model, the $\gamma_d$IDE model presents a relatively lower value of dust matter fraction in the universe ($\Omega_m=0.2690_{-0.0232}^{+0.0242}$). This result is in tension with recent constraint results on $\Omega_m$ based on Planck measurements of CMB: the best-fit value is $\Omega_\Lambda=0.315\pm 0.017$ in the flat case \cite{Ade13}.

\section{A quantitative criteria for coincidence problem}\label{sec5}

As mentioned in the introduction, the interacting dark energy models are often invoked to explain and test the coincidence problem.
Now we introduce the ratio $r$ between $\rho_m$ and $\rho_X$ so that $r=\rho_m/\rho_X$. If the ratio $r=\rho_m/\rho_X$ does not change much during the whole history of the universe, then the coincidence problem can be solved. By combining Eq.~(\ref{eq4}), we obtain the parameter $r$ evolving as
\begin{equation}
\dot{r}=3Hr(1+r)[\frac{w_X}{1+r}+\frac{Q}{3H\rho_m}]
\label{attractor}
\end{equation}
Moreover, the attractor solution in the interacting model is a necessary ingredient to alleviate the coincidence problem. From Eq.~(\ref{attractor}), we find the attractor solutions for both IDE models depend on the values of interaction term and the EoS parameter of dark energy. However, compared with the $\gamma_m$IDE model, the negative interaction term in the $\gamma_d$IDE model leads to negative $r_s$, which is not physical.

\begin{center}
 \begin{figure}[htbp]
 \centering
 \includegraphics[width=0.5\textwidth]{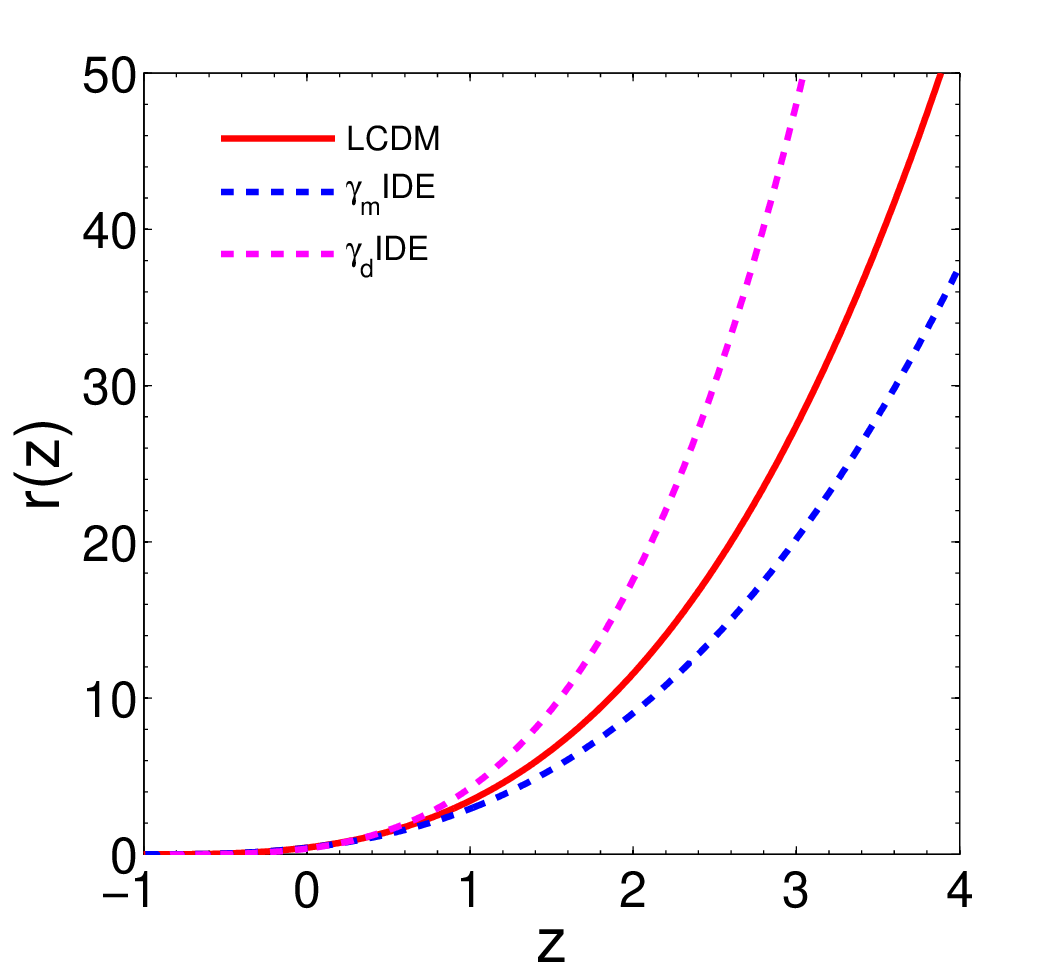}\includegraphics[width=0.5\textwidth]{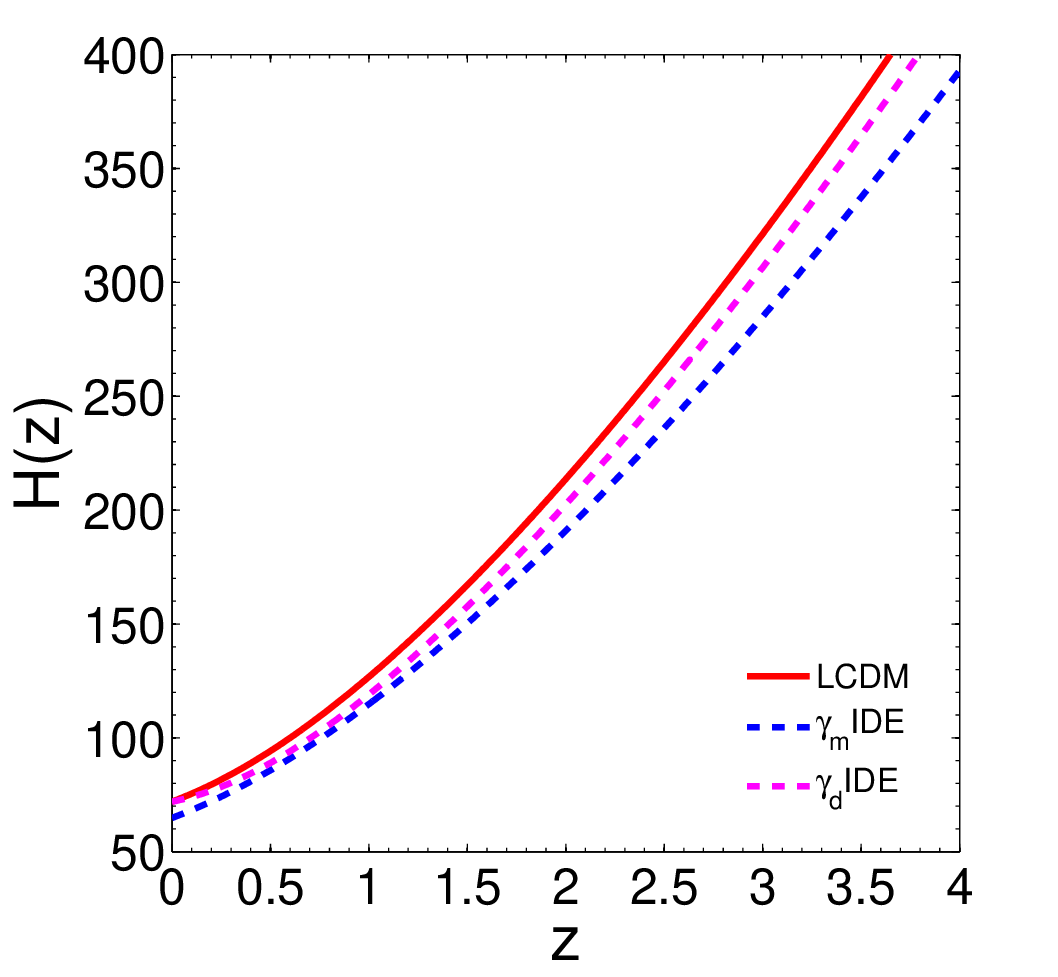}
 \caption{\label{evolution} The evolution of the ratio $r=\rho_m/\rho_X$ as a function of the redshift (Left) and $H$ as a function of z in $\Lambda$CDM, $\gamma_m$IDE, and $\gamma_d$IDE models (Right). The model parameters are taken as the best-fit values from OHD+SNe+BAO+CMB.}
 \end{figure}
 \end{center}

Fig.~\ref{evolution} illustrates other constraint results from the joint analysis of OHD, SNe Ia, BAO and CMB. The
left and right panel display the evolutions of the ratio $r$ and Hubble parameter $H(z)$ as a function of the redshift, respectively.
On the one hand, it indicates that the densities of dark energy and dark matter are the same order of magnitude in the redshift range $0.0\leq z \leq 2.0$.
On the other hand, the positive coupling obtained from the $\gamma_m$IDE model seems to lead to a slower change of $r$ as compared to the noninteracting case in the $\Lambda$CDM model and the negative coupling in the $\gamma_d$IDE model. Therefore, the coincidence problem is less acute concerning this interacting dark energy model.

Based on these considerations, we choose to apply a quantitative criteria to judge the severity of the coincidence problem \cite{Zhang09}. Under this criteria, two indices concerning early coincidence ($C_e$) and late coincidence ($C_f$) are defined
\bea
 &&C_e=\frac{r_e}{r_0}, \nonumber\\
 &&C_f=\frac{r_f}{r_0},\label{eq4}
 \eea
where $r_e$, $r_0$, and $r_f$ represent the energy ratio at early time, at present, and at the attractor value (if it exists).
In order to determine the value of $C_e$, we take $z=100$ as a standard epoch in the early universe \cite{Zhang09}. Based on the calculation results of the two indices, we also include the third index of coincidence $C$ in the whole history of the universe
\bea
 &&C=F(C_e)F(C_f),
 \label{eq4}
 \eea
where $F(x)$ is a function defined to avoid the problem that $C_e$ and $C_f$ might be vary in the opposite direction \cite{Zhang09}.

We apply this criteria to the two interacting models and the results are shown in Table~\ref{indices}. For comparison, the indices for other two IDE models, the interacting quintessence and phantom models obtained by \cite{Zhang09} are also listed. Let us briefly comment the results in the presence of different forms of coupling. As can be seen in Table~\ref{indices}, the coincidence index $C$ for the $\gamma_m$IDE model is smaller than that for the $\gamma_d$IDE model, the interacting quintessence and phantom models by four orders of magnitude, which provides a piece of positive evidence that the $\gamma_m$IDE model may alleviates the coincidence problem.

\begin{table}[ph]
\tbl{Coincidence indices for the $\gamma_m$IDE model and $\gamma_d$IDE model. For comparison, the indices for interacting quintessence and phantom models are also listed.}
{\begin{tabular}{c|c|c|c|c}\hline\hline
 Indices & \hspace{4mm} $\gamma_m$IDE & \hspace{4mm} $\gamma_d$IDE & \hspace{4mm} IQT \cite{Zhang09} & \hspace{4mm} IPT \cite{Zhang09}  \\ \hline
 $C_e$ & $7.4\times 10^3$ &  $1.0\times10^7$ &  $1.0\times10^4$ &  $1.0\times10^4$ \\
 $C_f$ & $5.0$ &  $-$  & $1.0\times 10^4$ & $1.0\times 10^4$ \\
 $C$   & $3.7\times 10^4$ &  $>1.0\times 10^7$  & $1.0\times 10^8$ & $1.0\times 10^8$ \\
 \hline\hline

\end{tabular} \label{indices}}
\end{table}

\section{Conclusions}\label{sec6}

In this paper, we have examined, with the newly revised OHD
versus redshift data, the cosmic microwave background (CMB)
detected by the 9-year WMAP data, the baryonic acoustic oscillation
(BAO) peak detected by large-scale correlation function of luminous
red galaxies from Sloan Digital Sky Survey (SDSS) data release 7 (DR7), SDSS-III Baryon Oscillation Spectroscopic Survey (BOSS), WiggleZ survey, and 6dFGS survey, as well as the newly revised Union2 SNe Ia data set, to constrain two phenomenological interaction models for dark energy and dark matter, which are proposed as
candidates to ease the coincidence problem of the concordance $\Lambda$CDM model. We find that, for the $\gamma_m$ IDE and
$\gamma_d$ IDE models where $\gamma_m$ and $\gamma_d$ quantify the
extent of interaction, although the OHD data can not tightly
constrain the model parameters, stringent constraints can be
obtained in combination with the CMB, BAO, and Union2 SNe Ia observations.

For the $\gamma_m$ IDE model, we obtain the best-fit values for the parameters: $(w_X, \gamma_m, \Omega_m)=(-0.9378_{-0.1619}^{+0.1726},0.0073_{-0.0034}^{+0.0048},0.2940_{-0.0268}^{+0.0287})$. For the $\gamma_d$ IDE model, the best-fit values are $(w_X, \gamma_d, \Omega_m)=(-1.1502_{-0.1641}^{+0.1521},-0.0137_{-0.0153}^{+0.0147},0.2690_{-0.0232}^{+0.0242})$. These results are consistent with and more stringent than the previous constraint analysis \cite{Guo07,Feng08,Chen10}. More precisely, we find the same tendencies for both phenomenological forms of interaction term $Q$, i.e., the parameter $\gamma$ to be a small number, $|\gamma|\approx 10^{-2}$. The small coupling result is consistent with that obtained independently by galaxy cluster analysis \cite{Abdalla07} and combined analysis including the look back time data \cite{Feng08}. However, the difference between the constraint results made with the two interacting dark energy models is also most sharpest concerning the sign of the interaction parameter $\gamma$. We observe that $\gamma>0$ when the interaction between dark sectors proportional to the energy density of dust matter, whereas the negative coupling ($\gamma<0$) is preferred by the latest observations when the interaction term is proportional to the DE energy density. The positive coupling is required to alleviate the coincidence problem, which is also the requirement of the second law of thermodynamics.

We also find two possible explanations to this incompatibility. One is the proportional relation between the values of the EoS parameter and the interaction term $\gamma_d$ ($\gamma_m$), which can be seen from Fig.~\ref{1.1}-\ref{2.1}. Besides, the actual values of the equations of state of dark energy are quite different for the two phenomenological forms of interaction. The $\gamma_d$IDE model seems to behave in the same way as a quintom dark energy model. However, the $\gamma_m$IDE model exhibits a quintessence behavior and it has more possibility for the EoS parameter to cross quintessence divide line $w_X=-1$ at $1\sigma$ C.L. The other possible explanation of this discrepancy is the value of $\Omega_m$, compared with the $\gamma_m$IDE model, the $\gamma_d$IDE model presents a relatively lower value of dust matter fraction in the universe, which is in tension with recent constraint results on $\Omega_m$ based on Planck measurements.

Finally, in order to study the coincidence problem through a detailed analysis, we also apply a quantitative criteria for the determination of the severity of the coincidence problem. Applying this criteria to the $\gamma_m$IDE and $\gamma_d$IDE models, we find that the $\gamma_m$IDE model may alleviate the coincidence problem, because its coincidence index $C$ is smaller than that for the $\gamma_d$IDE model, the interacting quintessence and phantom models by four orders of magnitude. We are looking forwards to see whether these results may be changed with future observational data including high
redshift SNe Ia data from SNAP \cite{Abraham04}, more precise Gamma Ray Bursts data
\cite{Schaefer07,Basilakos08}, and gravitational lensing data \cite{Zhu98,Sereno02,Cao12a,Cao12b}.

\section*{Acknowledgments}

This work is supported by the National Natural Science
Foundation of China under the Distinguished Young Scholar
program (Grant Nos. 10825313 and 11073005), the National Basic Research Program of China (973
program, Grant No. 2012CB821804), and the Fundamental Research Funds for the Central Universities
and Scientific Research Foundation of Beijing Normal University.

\end{document}